\documentclass[11pt]{article} \usepackage{mystyle}
\usepackage{epsfig,amsmath} \usepackage{graphics}
\usepackage{cite,./mcite}
 
  \newenvironment{defl}[1]%
  {\begin{list}{}{\settowidth{\labelwidth}{#1}%
  \setlength{\leftmargin}{\labelwidth}%
  \addtolength{\leftmargin}{\labelsep}%
  \setlength{\itemsep}{0pt plus 1pt}
  \setlength{\parsep}{0pt plus 1pt}
  \setlength{\topsep}{0pt plus 1pt}
  \setlength{\partopsep}{0pt plus 1pt}
  \setlength{\parskip}{2mm plus 1mm minus 1mm}
  }}%
  {\end{list}}
\def\lsim{\mathrel{\rlap{\lower4pt\hbox{\hskip1pt$\sim$}}
    \raise1pt\hbox{$<$}}}                
\def\gsim{\mathrel{\rlap{\lower4pt\hbox{\hskip1pt$\sim$}}
    \raise1pt\hbox{$>$}}}                
\newcommand{\cascadeversion}{~2.2.0}
\newcommand{\cA}{{\cal A}}
\newcommand{\as}{\alpha_\mathrm{s}}
\newcommand{\asb}{{\bar \alpha}_\mathrm{s}}

\newcommand{\Pmax}{\bar{q}}
\def\prp{t}
\newcommand{\kt}{k_{t}}
\def\kti#1{\ensuremath{k_{\prp #1}}}
\def\pti#1{\ensuremath{p_{\prp #1}}}
\newcommand{\pt}{p_{t}}

\newcommand{\SMALLXC}{SMALLXa,SMALLXb}
\newcommand{\CCFM}{CCFMa,CCFMb,Catani:1989sg,CCFMd}

\newcommand{\alphasb}{\bar{\alpha}_s}
\newcommand{\JETSETMC}{\PYTHIAMC}
\newcommand{\LEPTOMC}{Ingelman_LEPTO65}
\newcommand{\PYTHIAMC}{Pythia61}

\def\CASCADE{{\sc Cascade}}
\def\SMALLX{{\sc Smallx}}

\def\LEPTO{{\sc Lepto}}
\def\PYTHIA{{\sc Pythia}}
\def\HERWIG{{\sc Herwig}}
\def\JETSET{{\sc Jetset}}

\makeindex
\begin{document}
\begin{flushright}
DESY 10-107 \\
 August 2010
\end{flushright}
\begin{center} {\sffamily\Large\bfseries 
The CCFM Monte Carlo generator CASCADE \\ \vspace*{0.15cm}
Version\cascadeversion\ }
 \\ \vspace{0.5cm}
{ \Large H.~Jung$^{1,2}$,
S.~Baranov$^3$, M. Deak$^4$, A. Grebenyuk$^1$, F.~Hautmann$^5$,
M.~Hentschinski$^1$,  A.~Knutsson$^1$, M.~Kr\"amer$^1$, K.~Kutak$^2$, A.~Lipatov$^6$, N.~Zotov$^6$ }\\
      {\large $^1$DESY, Hamburg, FRG}\\
     {\large $^2$University of Antwerp, Antwerp, Belgium}\\
   {\large $^3 $Lebedev Physics Institute, Russia }\\
    {\large $^4$Instituto de F\'isica Te\'orica UAM/CSIC, University of Madrid, Spain }\\
    {\large $^5$University of Oxford, GB}\\
    {\large $^6$SINP, Moscow State University, Russia }
\end{center}
\begin{abstract}
\CASCADE\ is a full hadron level Monte Carlo event generator for $ep$, 
$\gamma p$ and $p\bar{p}$ and $pp$ processes,
 which uses the
CCFM evolution equation for the initial state cascade in a backward evolution
approach supplemented 
with off - shell matrix elements for the hard scattering.
A detailed program description is given, with emphasis on parameters
the user wants to change and common block variables which 
completely specify the generated events.
\end{abstract} 
{\sffamily\large\bfseries PROGRAM SUMMARY} \\ \\
{\em Title of Program:} \CASCADE\ \cascadeversion\ \\ \\
{\em Computer for which the program is designed and others on which it is
operable:}   any with standard Fortran 77 (g77 or gfortran), tested on 
                 SGI, HP-UX, SUN, PC, MAC\\ \\
{\em Programming Language used:}  FORTRAN 77 \\ \\
{\em High-speed storage required:}  No \\ \\
{\em Separate documentation available: } No 
{\em Nature of physical problem:} 
High-energy collisions of particles at moderate values of the fractional momentum $x$ are well 
described by resummation of leading logarithms of transverse momenta 
$(\as \ln Q^2)^n$, generally referred to as DGLAP physics. At small $x$ 
leading-logs of longitudinal momenta, $(\as \ln x)^n$, are expected to become
equally if not more important (BFKL). An appropriate description valid for
both small and moderate $x$ is given by the CCFM evolution equation, resulting 
in an unintegrated gluon density  ${\cal A} (x,\kt,\Pmax ) $, which is also a 
function of the evolution scale~$\Pmax $. 
\\ \\
{\em Method of solution:}  
Since measurements involve complex cuts and multi-particle final states, the 
ideal tool for any theoretical description of the data is a Monte Carlo 
event generator which embodies small-$x$ resummation, in analogy to event 
generators which use DGLAP resummation. The CCFM evolution equation forms a bridge between the DGLAP and BFKL resummation and can be applied to generate the initial state branching processes. The CCFM equation  can be formulated in a way suitable for carrying out a 
backward evolution, which is an essential requirement to efficiently 
generate unweighted Monte Carlo events.
\\ \\
{\em Restrictions on the complexity of the problem:}  
The following hard subprocesses can be simulated:
$\gamma^* g^* \rightarrow q \bar{q} (Q\bar{Q})$, 
$\gamma g^* \rightarrow J/\psi g $, 
$ g^* g^* \rightarrow q \bar{q} (Q\bar{Q})$,
$ g^* g^* \to J/\psi g $, 
$ g^* g^* \to \chi $,
$g^*g \to g g$, 
$g^* q \to g q$, 
$ g^* g^* \rightarrow h^0$,
$g^* q \to Z(W)  q$, 
$g^* g^* \to Z Q \bar{Q} $, $g^* g^* \to Z q \bar{q} $, 
$g^* g^* \to W q_i q_j$.

The
present version is applicable for HERA, TEVATRON and LHC processes. \\ \\
{\em Other Program used:}  \PYTHIA\ ({\it version $>$ 6.4}) for hadronisation, 
{\sc Bases/Spring}  5.1 
for integration (supplied with the program package).\\ \\ 
{\em Unusual features of the program:}   None \\ \\

\newpage

\section{The CCFM evolution equation}
\label{sec:CCFMEquation}

The formulation of the CCFM~\cite{\CCFM} parton evolution 
for the implementation into a full hadron level Monte Carlo program is
described in detail
in~\cite{CASCADE,jung_salam_2000}.  Here only 
the main results are summarized and discussed.
\begin{figure}[th]
\centerline{\resizebox{0.9\textwidth}{!}{\includegraphics{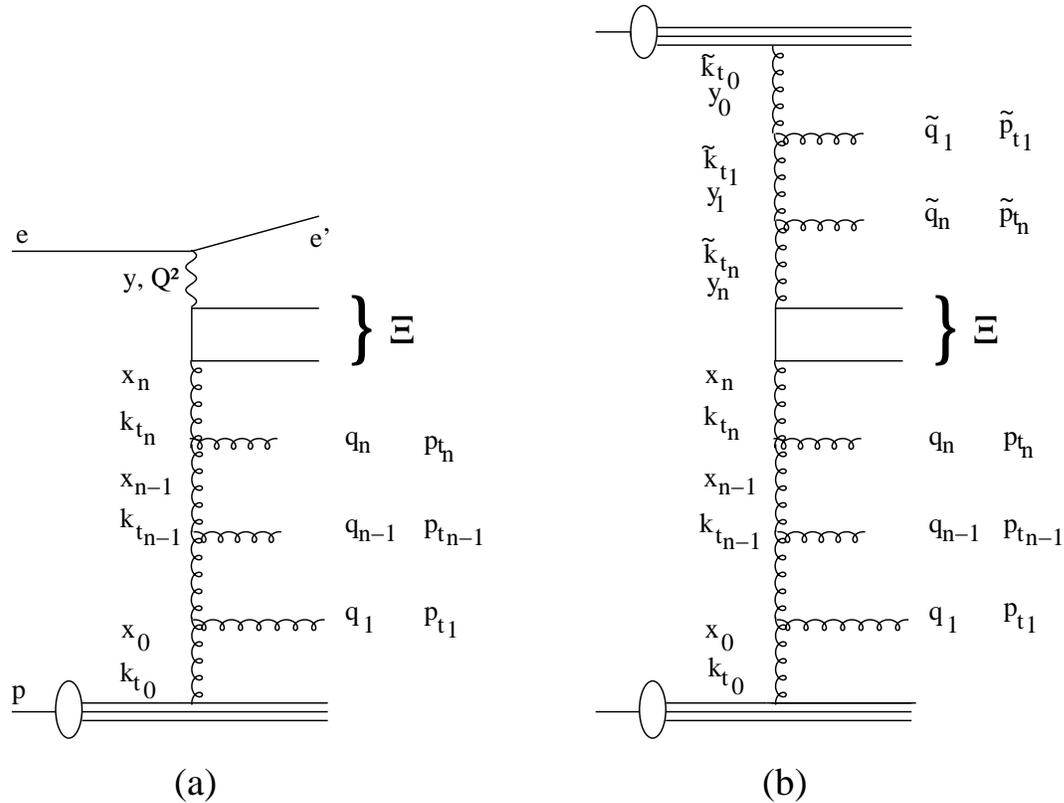}}} 
\vspace*{8pt}
\caption{\it Kinematic variables for multi-gluon emission in leptoproduction ($a$)
and hadroproduction ($b$). 
The $t$-channel gluon
four-vectors are given by $k_n$ and the gluons emitted in the initial state
cascade have four-vectors $p_n$. The maximum
angle (a function of the rapidity) for any emission is obtained
from the quark box, as indicated with $\Xi$.
\label{CCFM_variables} } 
\end{figure}
The pattern of QCD initial state
radiation in a small-$x$ event in $ep$ and $p \bar{p}$($pp$)  collisions
 is illustrated in
Fig.~\ref{CCFM_variables}  together with labels for the
kinematics.  According to the CCFM evolution equation, the emission of
partons during the initial cascade is only allowed in an
angular ordered region of phase space. 
 In terms
of Sudakov variables $\Upsilon$ and $\Xi$  the quark pair momentum is written as:
\begin{equation}
p_q + p_{\bar{q}} = \Upsilon (p^{(1)} + \Xi p^{(2)}) + Q_t \;,
\end{equation}
where $p^{(1)}$ and $p^{(2)}$ are the four-vectors of
 incoming particles (electron-proton, proton-antiproton or proton-proton), 
respectively and $Q_t$ is the transverse momentum of the quark pair
in the center of mass frame of
$p^{(1)}$ and $p^{(2)}$ (cms).
The variable $\Xi$ is related to the rapidity $Y$ 
in the center of mass (CMS) frame via 
\begin{equation}
Y = \frac{1}{2} \log \left(\frac{E+p_z}{E-p_z}\right)  
= \frac{1}{2}  \log \left(\frac{1}{\Xi}\right)  .
\end{equation}
Using $E=E_q+E_{\bar{q}}$ and $p_z = p_{q\;z} + p_{\bar{q}\;z}$ gives 
   $E+p_z= \Upsilon \sqrt{s}$, $E-p_z= \Upsilon \Xi\sqrt{s}$ with
$E=\sqrt{s}/2$ and 
$s=(p^{(1)}+p^{(2)})^2$ being the squared center of mass energy.
 Therefore $\Xi$ can be used to define the maximum allowed angle 
in the evolution.
The momenta $p_i$ of the gluons emitted during the initial
state cascade are given by (here treated massless):
\begin{equation}
p_i = \upsilon_i (p^{(1)} + \xi_i p^{(2)}) + \pti{i} \;  , \;\; 
\xi_i=\frac{p_{ti}^2}{s \upsilon_i^2},
\end{equation}
with $\upsilon_i = (1 - z_i) x_{i-1}$ and $x_i = z_i x_{i-1}$. 
The variables 
 $x_i$ and $\upsilon_i$ are the
momentum fractions of the exchanged and emitted gluons, while $z_i$ is
the momentum fraction in the branching $(i-1) \to i$ and $\pti{i}$ is
the transverse momentum of the emitted gluon $i$. 
Again the rapidities $y_i$ are given by $y_i = - 0.5 \log \xi_i $ in the CMS frame. 
\par
The angular ordered region is then specified by 
(Fig.~\ref{CCFM_variables}$a$ and the lower part of the
cascade in Fig.~\ref{CCFM_variables}$b$, for the upper part  the variables
have to be changed accordingly):
\begin{equation}
\xi_0 < \xi_1< \cdots < \xi_n < \Xi ,
\end{equation}
which becomes:
\begin{equation}
z_{i-1} q_{i-1} < q_{i}  ,
\end{equation}
where the rescaled transverse momentum $q_{i}$ of the emitted
gluon is defined by:
\begin{equation}
 q_{i} = x_{i-1}\sqrt{s \xi_i} = \frac{\pti{i}}{1-z_i} .
 \label{qbar}
\end{equation}
\par 
The scale $\Pmax$ (related to the maximum angle) 
can be written as:
\begin{equation}
 \Pmax^2 = \Upsilon^2 \Xi s 
 = \hat{s} + Q_{\prp}^2  ,
\end{equation}
with $\hat{s}=(p_q + p_{\bar{q}})^2$ and the relation of $\Pmax$
to a particular choice of the factorisation scale 
$\mu_f$ in the collinear approach becomes obvious.
\par
The CCFM evolution equation can be written in a differential form~\cite{CCFMd},
which is best suited for a backward evolution approach adopted in the Monte
Carlo generator \CASCADE\ ~\cite{CASCADE,jung_salam_2000}:
\begin{equation}
\Pmax^2\frac{d\; }{d \Pmax^2} 
   \frac{x \cA(x,\kt,\Pmax)}{\Delta_s(\Pmax,Q_0)}=
   \int dz \frac{d\phi}{2\pi}\,
   \frac{\tilde{P} (z,\Pmax/z,\kt)}{\Delta_s(\Pmax,Q_0)}\,
 x'\cA(x',\kt',\Pmax/z) ,
\label{CCFM_differential}
\end{equation} 
where $\cA(x,\kt,\Pmax)$ is the unintegrated gluon density, depending on 
$x$, $\kt$ and the evolution variable $\Pmax$. The splitting variable is 
$z=x/x'$ and $\vec{\kt}' = (1-z)/z\vec{q} + \vec{\kt}$, where the vector
$\vec{q}$ is at an azimuthal angle $\phi$.
The  Sudakov form factor $\Delta_s$ is given by:
\begin{equation}
\Delta_s(\Pmax,Q_0) =\exp{\left(
 - \int_{Q_0^2} ^{\Pmax^2}
 \frac{d q^{2}}{q^{2}} 
 \int_0^{1-Q_0/q} dz \frac{\alphasb(q^2(1-z)^2)}{1-z}
  \right)} ,
  \label{Sudakov}
\end{equation}
with $\alphasb=\frac{C_A \alpha_s}{\pi}=\frac{3 \alpha_s}{\pi}$. For
inclusive quantities at leading-logarithmic order the Sudakov form
factor cancels against the $1/(1-z)$ collinear singularity of the
splitting function. 
\par
The original splitting function $\tilde{P}_g (z_i,q_i,k_{ti})$ for branching $i$ 
is given by (neglecting finite terms as they
are not obtained in CCFM at the leading infrared
accuracy~(cf p. 72 in \cite{Catani:1989sg}):
\begin{equation}
\tilde{P}_g (z_i,q_i,k_{ti})
= \frac{\alphasb(q^2_{i}(1-z_i)^2)}{1-z_i} + 
\frac{\alphasb(k^2_{ti})}{z_i} \Delta_{ns}(z_i,q^2_{i},k^2_{ti}),
\label{Pgg}
\end{equation}
where the non-Sudakov form factor $\Delta_{ns}$ is defined as:
\begin{equation}
\log\Delta_{ns} =  -\alphasb(k^2_{ti})
                  \int_0^1 \frac{dz'}{z'} 
                        \int \frac{d q^2}{q^2} 
              \Theta(k_{ti}-q)\Theta(q-z'q_{ti}).
                  \label{non_sudakov}                   
\end{equation}
\par
The implementation of the full splitting function including non singular terms can lead to inconsistencies.
Replacing naively only
$  \frac1{1-z} \to \frac{1}{1-z} - 2 + z(1-z)\,$ 
in the CCFM splitting function can lead to negative
branching probabilities. 
\par
In \cite{smallx:2001} it was suggested to use:
\vspace*{-0.2cm}
\begin{eqnarray}
P(z,q,k)& = &\asb \left(\kt^2\right) \left( \frac{(1-z)}{z} + (1-B)z(1-z)\right)
\Delta_{ns}(z,q,k) \label{Pgg_fullsplitt}
\\
& &  + \asb\left((1-z)^2q^2\right) \left(\frac{z}{1-z} + Bz(1-z)\right),
\nonumber
\end{eqnarray}
where $B$ is a parameter to be chosen arbitrarily between $0$ and $1$,
we take $B=0.5$. 
As a consequence of the replacement, the Sudakov form factor will change, but 
also the non-Sudakov form factor needs to be replaced by:
\begin{equation}
\log\Delta_{ns} =  -{\bar \alpha}_s\left(\kt^2\right)
                  \int_0^1 dz'
                        \left( \frac{1-z}{z'} + (1-B)z(1-z) \right)
                        \int \frac{d {q'}^2}{{q'}^2} 
              \Theta(k-q')\Theta(q'-z'q).
\label{non_sudakov_fullsplitt}
\end{equation}

\section{Backward evolution: CCFM and CASCADE }
\label{sec:Backward}

The idea of a backward evolution~\cite{PYTHIAPSa,LEPTOPS}
is to first generate the hard
scattering process with the initial parton momenta distributed
according to the parton distribution functions.  This involves in
general only a fixed number of degrees of freedom, and the hard
scattering process can be generated quite efficiently. The
initial state cascade is generated by going backwards from the hard
scattering process towards the beam particles. 

According to the CCFM equation the probability of finding a gluon in
the proton depends on three variables, the momentum fraction $x$, the
transverse momentum squared $k_t^2$ of the exchanged gluons and the
scale $\Pmax = x_{n} \sqrt{s \Xi}$, which is related to the
maximum angle $\Xi$ allowed for any emission. 
To solve eq.(\ref{CCFM_differential}) the unintegrated parton
distribution $\cA(x,\kt,\Pmax)$ has to be determined beforehand. 
\par
Given $\cA(x,\kt,\Pmax)$, the generation of a full hadronic event is separated
into three steps, as implemented in the hadron level Monte Carlo program 
\CASCADE: 
\begin{itemize}
\item[$\bullet$] 
The hard scattering process is generated,
\begin{equation}
\sigma = \int d\kti{1}^2 d\kti{2}^2 dx_1dx_2 {\cal A}(x_1,\kti{1},\Pmax)
{\cal A}(x_2,\kti{2},\Pmax) \hat{\sigma} (k_1+k_2 \to X )\,,
\label{x_section}
\end{equation}
with $k_1 (k_2)$ being the momenta of the incoming partons to the subprocess $k_1 + k_2 \to X$ with  $X$ being the final state. The definition of $\hat{\sigma}$ follows~\cite{Catani:1990eg}. 
The available processes are shown in Tab~\ref{processes}.
The momenta of the incoming partons  are  given in Sudakov representation:
\begin{eqnarray*}
 k_1 = x_1 p^{(1)} + \bar{x}_2 p^{(2)} + k_t \simeq x_1 \,p^{(1)}  + \kti{1}\\
 k_2 = \bar{x}_1 p^{(1)} + x_2 p^{(2)} + k_t \simeq x_2 \,p^{(2)}  + \kti{2}
 \label{xgluon}
\end{eqnarray*}
where the last expression comes from the high energy approximation ($x_{1,\;2} \ll
1$), which then gives  $-k^2 \simeq k_t^2$.
\item[$\bullet$] The initial state cascade is generated according to
  CCFM in a backward evolution approach.
\item[$\bullet$] The hadronisation is performed using the Lund string
  fragmentation implemented in \PYTHIA\ /\JETSET \cite{\JETSETMC}.
\end{itemize}

The parton virtuality enters the hard scattering process and also influences
the kinematics of the produced particles ($Z_0,\, W,$ Higgs and quarks) and therefore the maximum angle
allowed for any further emission in the initial state cascade. This
virtuality is only known after the whole cascade has been generated,
since it depends on the history of the parton evolution
(as $\bar{x}$ in eq.(\ref{xgluon}) may not be neglected for exact
kinematics).  In the
evolution equations itself it does not enter, since there only the
longitudinal energy fractions $z$ and the transverse momenta are
involved.  This problem can only approximately be overcome by using
$k^2 = k_t^2/(1-x)$ for the virtuality which is correct in the case
of no further parton emission in the initial state.

The Monte Carlo program \CASCADE\ can be used to generate unweighted
full hadron level events, including initial state parton evolution
according to the CCFM equation and the off-shell matrix elements for
the hard scattering process. It is applicable for $p \bar{p}$, $pp$, 
photoproduction as well as for deep inelastic scattering.
A discussion of the phenomenological applications of \CASCADE\  can be found in \cite{Hautmann:2008vd}.

The typical
time needed to generate one event  is
similar to the time needed by standard Monte Carlo event generators
like \PYTHIA~\cite{\PYTHIAMC}.

\subsection{The unintegrated parton density}

\begin{table}[htdp]
\begin{center}
\begin{tabular}{|c|l|c|c|c|c|c|c|}
\hline 
parton & uPDF set & \multicolumn{2}{|c|}{$x{\cal A}_0(x,k_t,\Pmax)$}& $\Lambda^{(4)}_{qcd}$ & $\kt^{cut}$ & $Q_0$ &ref \\ 
& & \multicolumn{2}{|c|}{$= Nx^{-B}(1-x)^C$}& & & & \\
\hline
 & & $B$ & $C$ & & & & \\
\hline
gluon  & set JS  &0 &4& 0.25 & 0.25 & 1.4 & \protect\cite{jung-dis04} \\
          & set A0  &0 &4& 0.25 & 1.3 & 1.3 & \protect\cite{jung-dis04} \\
          & set A0+  & -0.01&4&0.25  & 1.3  & 1.3 &  \protect\cite{jung-dis04} \\
          & set A0- & -0.01&4&0.25  & 1.3  & 1.3 &  \protect\cite{jung-dis04}\\ 
          & set A1 & -0.1&4&0.25  & 1.3  & 1.3 &  \protect\cite{jung-dis04}\\ 
          & set B0 & 0&4& 0.25 &  0.25 & 1.3 &  \protect\cite{jung-dis04}\\
         & set B0+&0.01&4 &0.25  & 0.25  &1.3 &  \protect\cite{jung-dis04}\\
         & set B0-&  0.01&4&0.25  & 0.25 & 1.3 & \protect\cite{jung-dis04}\\ 
         & set B1& -0.1 &4&0.25  & 0.25 & 1.3 & \protect\cite{jung-dis04}\\ 
 & set C & 0.25&4& 0.13 & 1.1 & 1.1 &  \protect\cite{jung-dis07}\\  
 & set 1 & 0& 4& 0.25 & 1.33&1.33 & \protect\cite{jung-dis03}\\  
 & set 2 & 0&4& 0.25 & 1.18 & 1.18&  \protect\cite{jung-dis03}\\  
 & set 3&0& 4& 0.25 & 1.35&1.35 & \protect\cite{jung-dis03}\\  
\hline
quark  & set A  & -- &--& 0.25 & 1.3  & 1.3 &  \\
\hline
\end{tabular}
\caption{\it Recommended CCFM unintegrated parton distribution functions included in \protect\CASCADE . See also sec.~\protect\ref{sec:alphas}.}
\end{center}
\label{updfs}
\end{table}%

The CCFM unintegrated parton density $x {\cal A}(x,k_{t},\Pmax)$ can be
obtained from a forward evolution procedure as implemented in
\SMALLX~\cite{\SMALLXC} by a fit to the measured structure function $F_2$ as
described i.e. in~\cite{CASCADE,jung_salam_2000}. 
From the initial parton distribution, which includes a Gaussian intrinsic $\kt$ distribution,
 a set of values $x$
and $k_{t}$ is obtained by evolving up to a given scale $\log
\Pmax$ using a forward evolution procedure. 
Technically the parton density is stored on a grid in $\log x$, $\log k_{t}$ and $\log \Pmax$ and a linear interpolation is used to obtain the parton density for values in between the grid points.
The data file (i.e. \verb+ccfm-xxxx.dat+) containing the  
grid points 
is read in at the beginning of the program.
\par
Several sets ({\bf J2003 set 1 - 3}~\cite{jung-dis03} (
\verb+IGLU=1001-1003+), 
{\bf set A}~\cite{jung-dis04} \verb+IGLU=1010-1013+ and 
{\bf set B}~\cite{jung-dis04} \verb+IGLU=1020-1023+) of
unintegrated gluon densities are available with 
the input parameters  fitted to
describe the structure function $F_2(x,Q^2)$ 
in the range $x < 5 \cdot 10^{-3}$ and  $Q^2 > 4.5$~GeV$^2$ 
as measured at H1~\cite{H1_F2_1996,H1_F2_2001} and  
ZEUS~\cite{ZEUS_F2_1996,ZEUS_F2_2001}. 
Set {\bf JS}~\cite{jung_salam_2000} (\verb+IGLU=1+) is fitted only to 
$F_2(x,Q^2)$ of Ref.~\cite{H1_F2_1996}.  
The collinear cutoff $\kt^{cut}=Q_0$ which regulates the region of  $z \to 1$, 
is applied both to the real emissions as well as inside the Sudakov form factor. 
 In {\bf JS}, {\bf J2003 set 1 - 3} and {\bf set A} we have 
$\kt^{cut}=Q_0=1.3$~GeV. 
Similarly, fits can be obtained using different values for the soft cut
$\kt^{cut}=0.25$~GeV, which are available in {\bf set B}. 
The {\bf set C} (\verb+IGLU=1101+)~\cite{jung-dis07} 
uses the full splitting function, as
described in eq.(\ref{Pgg_fullsplitt}), with a value for
$\Lambda^{(4)}_{QCD}=0.13$~GeV. This set was obtained by fitting simultaneously
the inclusive $F_2(x,Q^2)$ and jet measurements in DIS resulting in a changed 
 intrinsic $\kt$ distribution.
A CCFM parametrisation of the valence quark distribution is available (using as
starting distribution CTEQ 5 \cite{Lai:1999wy} and evolved with a splitting function
$P_{qq}$~\cite{Catani:1989sg} including angular ordering of the emitted gluon).
The available CCFM uPDF sets with the parameters of the starting distributions are listed in Tab.~\ref{updfs}.
\par
With the parameter \verb+IGLU+ also other unintegrated gluon densities
are accessible:
a simple numerical derivative of a standard integrated gluon density
$\frac{d xg(x,Q^2)}{dQ^2}$ taken
from~\cite{GRV95}  (\verb+IGLU=2+), the one in the
approach 
of Bl\"umlein~\cite{Bluemlein} and coded 
by~\cite{baranov_zotov_1999,baranov_zotov_2000} (\verb+IGLU=3+),
the unintegrated gluon density of KMS\footnote{A. Stasto kindly
provided the program code.}~\cite{martin_stasto} 
(\verb+IGLU=4+, stored in \verb+kms.dat+), 
the one of the saturation model by~\cite{wuesthoff_golec-biernat}
 (\verb+IGLU=5+, i.e. parameter set including charm)\footnote{The values of $\as$ and quark masses of \protect\cite{wuesthoff_golec-biernat} are not automatically used in the cross section calculation, but need to be set explicitly. } 
 and the one of KMR\footnote{M. Kimber kindly
provided the program code .}~\cite{martin_kimber} (\verb+IGLU=6+, 
stored in \verb+kmr.dat+). 

Initial state parton showers can be only generated for the CCFM unintegrated gluon density (with \verb+IGLU=1+ and 
\verb+IGLU=1001-1023,1101+). For all other sets only the cross section can be calculated without explicit inclusion of initial state parton showers, since the angular variable,
essential for angular ordering in the initial state cascade, is not available in the uPDFs.
However, the transverse momenta of the incoming partons are properly treated. 
Only the KMR set (\verb+IGLU=6+) provides a prescription for the emission of at
most one additional gluon.
\section{Hard processes in CASCADE}
Different sets of hard processes applicable for lepto (photo) - and hadroproduction have been calculated and are implemented in \CASCADE . The available processes are listed in Tab.~\ref{processes}.
\begin{table}[htdp]
\begin{center}
\begin{tabular}{|l||l|c|c|}
\hline 
Lepto(photo)production &process & IPRO & Reference \\ 
\hline
&$\gamma^* g^* \to q\bar{q} $ & 10 & \protect\cite{Catani:1990eg} \\
&$\gamma^* g^* \to Q\bar{Q} $ & 11 & \protect\cite{Catani:1990eg} \\ 
&$\gamma^* g^* \to J/\psi g $ & 2 & \protect\cite{saleev_zotov_a,Lipatov:2002tc, Baranov:2003at,Baranov:2002cf} \\
\hline
Hadroproduction& & &  \\
\hline
&$g^* g^* \to q\bar{q} $ & 10 & \protect\cite{Catani:1990eg} \\
&$g^* g^* \to Q\bar{Q} $ & 11 & \protect\cite{Catani:1990eg} \\ 
&$g^* g^* \to J/\psi g $ & 2 & \protect\cite{Baranov:2002cf} \\
&$g^* g^* \to \Upsilon g $ & 2 & \protect\cite{Baranov:2002cf} \\
&$g^* g^* \to \chi_c $ & 3 & \protect\cite{Baranov:2002cf} \\
&$g^* g^* \to \chi_b $ & 3 &\cite{Baranov:2002cf} \\
&$g^* g^* \to h^0 $ & 102 & \protect\cite{hautmann-higgs} \\
&$g^* g^* \to Z Q \bar{Q} $ & 504 & \protect\cite{Baranov:2008hj,Deak:2008ky} \\
&$g^* g^* \to Z q \bar{q} $ & 503 & \protect\cite{Baranov:2008hj,Deak:2008ky} \\
&$g^* g^* \to W q_i Q_j $ & 514 & \protect\cite{Baranov:2008hj,Deak:2008ky} \\
&$g^* g^* \to W q_i q_j $ & 513 & \protect\cite{Baranov:2008hj,Deak:2008ky} \\
&$q g^* \to Z q $ & 501 & \protect\cite{Marzani:2008uh} \\
&$q g^* \to qg $ & 10 & \protect\cite{deak} \\
&$g g^* \to gg $ & 10 & \protect\cite{deak} \\
\hline
\end{tabular}
\caption{\it Processes included in \protect\CASCADE . $Q$ stands for heavy quarks, $q$ for light quarks.}\label{processes}
\end{center}
\end{table}%
\subsection{Lepto(photo)production}  

\CASCADE\ can be used to simulate leptoproduction events over the whole $Q^2$
range. By fixing the light quark masses to  $m_q=0.25$ GeV
and $\alpha_s$ for small $\mu$, the hard scattering
matrix element remains finite over the full phase space.
The total cross section is simulated by selecting \verb+IPRO=10+ and 
\verb+NFLAV=4(5)+. With \verb+IPRO=10+ light quarks ($u,d,s$) are selected  and
with \verb+NFLAV>3+ the program automatically includes heavy flavour production
via the process \verb+IPRO=11+ and \verb+IHFLA=4+ up to \verb+IHFLA=NFLAV+.
The flag \verb+IRE1+ indicates, whether beam 1 has a internal structure: 
\verb+IRE1=1+ is used to generate resolved photon events. 
\par
Heavy flavour production can be generated separately via \verb+IPRO=11+. 
The value of \verb+IHFLA+ determines the heavy flavour to be generated.
\par
The matrix element for $\gamma g^* \to J/\psi (\Upsilon) g$ calculated in
\cite{Lipatov:2002tc,Baranov:2003at,Baranov:2002cf} is available for quasi-real $\gamma$'s via the process
\verb+IPRO=2+. The flavour of the Onium is selected via \verb+IHFLA+, i.e. \verb+IHFLA=4+ for $J/\psi$ and \verb+IHFLA=5+ for $\Upsilon$.
The matrixelement including $J/\psi (\Upsilon)$ polarisation and subsequent leptonic decay can be selected with \verb+IPSIPOL=1+.

\CASCADE\ can be used to simulate 
real photoproduction events by using \verb+KBE1=22+. The same options as for
leptoproduction are available. Resolved photon events can be generated with
\verb+IRE1=1+.

\subsection{Hadroproduction}
The hadroproduction processes available are listed in Tab~\ref{processes}.
The flavour code for beam 1 (2) can be
 chosen as \verb+KBE1=2212+ for proton or \verb+KBE1=-2212+ for anti - proton,
 for beam 2 \verb+KBE2+ is changed accordingly. 

\CASCADE\ can be used to simulate heavy quark production in
 $p p$ or $p \bar{p}$ collisions ($g^*g^* \to Q \bar{Q}$ 
 \verb+IPRO=11+ for heavy flavour production,
 and \verb+IHFLA=4(5)+ for charm (bottom) quarks), but also for light quarks with
 \verb+IPRO=10+. The matrix element for $g^* g^* \to J/\psi g$  calculated in
\cite{Lipatov:2002tc,Baranov:2003at,Baranov:2002cf} is available via the process
\verb+IPRO=2+. 
The matrixelement including $J/\psi$ polarisation and subsequent leptonic decay can be selected with \verb+IPSIPOL=1+. The process $g^*g^* \to \chi$ is available with \verb+IPRO=3+ including all three $\chi$ states with appropriate spin and angular momentum. The flavour of the Onium is selected via \verb+IHFLA+, i.e. \verb+IHFLA=4+ for $J/\psi (\chi_c)$ and \verb+IHFLA=5+ for $\Upsilon (\chi_b)$.

\par 
The process $g^*g^* \to  h^0 $ with the matrix element calculated 
in~\cite{hautmann-higgs} is
available via \verb+IPRO=102+, the Higgs mass can be selected via 
\verb+PMAS(25)+.

The process $g^*g^* \to  Z q\bar{q} $, calculated 
in~\cite{Baranov:2008hj,Deak:2008ky}, is
available via \verb"IPRO=503"  for light quarks and  \verb"IPRO=504"  for the heavy quarks. The flavor of the heavy quark is selected via \\
\verb"IHFLA=4 (5,6)"  for charm(bottom, top). 
The process $g^*g^* \to  W q_i q_j $, calculated 
in~\cite{Baranov:2008hj,Deak:2008ky}, is
available via \verb"IPRO=513" for light quarks and  \verb"IPRO=514" for heavy quarks, where the flavour of the heaviest quark is defined by \verb"IHFLA=3, (4,5)". Which of the quarks are produced depends on the charge of the $W$ which is randomly selected.

 The processes $g^*g \to g g$ and $g^*q \to gq$ \cite{deak} are included via process 
\verb+IPRO=10+. The individual processes can be selected for $g^*g^* \to q\bar{q}$  via
\verb+IRPA=1+, $g^*g \to gg$  via \verb+IRPB=1+ and $g^* q \to gq$  via \verb+IRPC=1+. Note that here one of the partons is treated on-shell. For the quarks the unintegrated quark distribution (for valence quarks) is used.

The process $qg^* \to Z q$, calculated in \cite{Marzani:2008uh,Baranov:2008hj}, is available via \verb"IPRO=501", using the unintegrated (valence) quark distribution for the on-shell quark.

\subsection{$\alpha_s$ and the choice of scales}
The strong coupling $\alpha_s$ is calculated via the \PYTHIA~\cite{\PYTHIAMC} 
subroutine \verb+PYALPS+. 
Maximal and minimal number of flavours used in $\alpha_s$ are set by
\verb+MSTU(113)+ and \verb+MSTU(114)+, $\Lambda_{QCD}$ = \verb+PARU(112)+ with respect to the number of flavours given in
\verb+MSTU(112)+ and stored in the \PYTHIA\ common block
\verb+COMMON/PYDAT1/+.
In the initial state cascade according to CCFM, the transverse momenta of the
$t$-channel gluons are allowed to perform a random walk for small $z$ values 
and $k_t$ can become very small. 
In the $1/z$ part of the splitting function we use $\mu=k_t$ as the scale in 
$\alpha_s(\mu)$ and in the $1/(1-z)$ part  $\mu=p_t$ is used. In
addition we require $\mu > Q_0$, resulting in 
$\alpha_s(\mu>Q_0) < 0.6$.
\par
The scale $\mu$, which is used in $\alpha_s$
in the hard scattering matrix element, can be changed
with the parameter \verb+IQ2+, the default choice is $\mu^2 = p_t^2$.
\par
The renormalisation scale dependence of the final 
cross section can be estimated by changing the scale used in $\alpha_s$ in the
off-shell matrix element. Since here we are using the LO $\alpha_s$ matrix
elements, any scale variation will change the cross section. In order to obtain
a reasonable result, the uPDF was fitted to describe $F_2$  by varying the scale
$\mu_r$. The {\bf set A0-,B0-} correspond to a scale $\mu_r = 0.5 \pt$ whereas
 {\bf set A0+,B0+}  correspond to a scale $\mu_r = 2 \pt$. 
\par
In order to investigate the uncertainties coming from the specific choice of the
evolution scale,  
another definition is applied, relating the factorisation scale only to the 
quark (or anti-quark):
$\mu_f  = \frac{\pt}{1-z}$
with $\pt$ being the transverse momentum of the quark (anti-quark) and 
$z = \frac{\tilde{k}_t}{y x_g s} $. The {\bf set A1,B1} 
correspond to a scale $\mu_f  = \frac{\pt}{1-z}$.
In the PDF {\bf set C} $\Lambda^{(4)}_{QCD}$ was fixed to
$\Lambda^{(4)}_{QCD}=0.13$~GeV.

\subsection{Quark masses}

The quark mass for light quarks ($u,d,s$)
 is fixed to $m_q=0.25$ GeV.
 This, together with the treatment of $\alpha_s$ at small scales
$\mu$, gives also a reasonable total cross-section for photoproduction at HERA
energies. The masses for heavy quarks are given by the \JETSET\ / \PYTHIA\ 
defaults ($m_c=1.5$ GeV, $m_b=4.8$ GeV) and can be changed according to the
\PYTHIA\ prescription.

\subsection{Initial and final state parton showers}

Initial state parton showers are generated in a backward evolution approach
described in detail in~\cite{CASCADE,jung_salam_2000}. The initial 
state parton shower
consists only of gluon branchings and is
generated in an angular ordered region in the laboratory frame. 
The gluons emitted in the initial state can undergo further timelike branchings. The maximum timelike mass $m_{max}$ is calculated using the angular constraint. With this mass, the parton which can further undergo a timelike branching is boosted to its rest frame with $m_{max}$ but keeping the original energy. The timelike branching is performed with the \PYTHIA\ routine \verb+PYSHOW+. After successful timelike branching, the proper mass is associated to the parton and the kinematics are calculated appropriately.
Gluon radiation from the valence quarks is also included.

\par
All parameters (like the scale $\mu$
in $\alpha_s$, the collinear cut-off $\kt^{cut}=Q_0$) for  the initial state cascade  
are fixed 
from the determination of the unintegrated gluon density. The
transverse momenta of the partons which enter the hard scattering 
matrix element
are already generated in the beginning and are not changed
during the whole initial and final state
parton showering.
\par
The final state parton shower uses the parton shower routine \verb+PYSHOW+ 
of \PYTHIA\
with the default scale $\mu^2=2\cdot (m_{1\;\perp}^2+m_{2\;\perp}^2)$ (\verb+IFIN=1+),
with $m_{1(2)\;\perp}$ being the transverse mass of the hard parton 1(2). Other choices are possible: $\mu^2=\hat{s}$ (\verb+IFIN=2+) and $\mu^2=2\cdot (m_1^2+m_2^2)$ (\verb+IFIN=3+). In addition a scale factor can be applied: \verb+SCAF+$\times\mu^2 $ (default:  \verb+SCAF=1+).

\subsection{Remnant treatment}

In \CASCADE\ version 1 the proton remnant was built in subroutine \verb+CAREMN+, which is a slightly
modified version of the \PYTHIA /\LEPTO\ subroutine \verb+PYREMN+. No intrinsic transverse momentum, in addition  to the transverse
momentum from the initial state cascade, was included.

From version 2.0 on the proton remnant can be generated directly via \PYTHIA\,
by selecting \verb+ILHA=10+ (which is now the default). 
The structure of the event record is then identical
to the one obtained from a standard \PYTHIA\ run. 

\subsection{Hadronization}

In \CASCADE\ version 1 the hadronization was done exclusively by \PYTHIA .  From version 2 onwards, events can be written into a file (via
switch \verb+ILHA=1+) according the LHA Accord \cite{lhaaccord},
which can be read by  any hadronization
program (like \PYTHIA\ or \HERWIG\ ), generating the remnants and performing the
hadronization.  With \verb+ILHA=10+   the hadronization is performed
within \PYTHIA .
The old \CASCADE\ format is obtained with \verb+ILHA=0+. Please note, that top decays can only be simulated properly within the \PYTHIA\ fragmentation and are therefore available only with  \verb+ILHA=10+.

\section{Description of the program components}
In \CASCADE\ all variables are declared as \verb"Double Precision". The 
Lund string model is used for hadronization as implemented in \PYTHIA\  
\cite{\PYTHIAMC}. The final state QCD radiation is performed via 
\verb"PYSHOW" from
 \PYTHIA\ . 
 The treatment of the proton remnant follows very closely the ones in
 \LEPTO \cite{\LEPTOMC} for the leptoproduction case and the one in 
\PYTHIA\  for the proton - proton case. However slight modifications were needed
to adapt to the cascade treatment here.
\par
The unintegrated gluon density is stored on data files
(\verb+ccfm.dat,kms.dat,kmr.dat+), 
and is read in at the beginning of the program.
\par
The program has to be compiled and linked together with \PYTHIA\ 6, to ensure that the double
precision code of \JETSET\  is loaded.
\subsection{Random number generator}
Since the variables are declared as double precision, also a double precision
random number generator has to be used to avoid any bias. The function
\verb+DCASRN+ gives a single random number,  the function \verb+DCASRNV+ returns
an array of length \verb+LEN+ of random numbers. The default random number
generator is \verb+RANLUX+ (called in \verb+DCASRN+ and \verb+DCASRNV+)
The source code of \verb+RANLUX+  (extracted from  {\sc Cernlib}) is included in the distribution. 
The user can change this to any preferred \verb+Double Precision+
random number generator.
\subsection{Integration and event generation}
The integration of the total cross section and the generation of unweighted
events is performed with the help of {\sc Bases/Spring} \cite{bases}, which is
included in the distribution package.

\subsection{Program history}
\begin{tiny}
\begin{verbatim}
CASCADE 
*________________________________________________________________________
*            Version  2.2.00  (Aug 2010)
*            timelike showering in initial state cascade 
*            forward DY process added
*            J/psi and chi_c (b) production added
*            all commons/variables now in double precision (version before cause
*            problems for LHC energies and small x)
*            obsolete routines from bases removed
*	       Bug in kinematics for Higgs prod corrected.
*            
*________________________________________________________________________
*            Version  2.1.00  (31. Dec 2009)
*            pt cuts are now done all in meoffsh routines on partons entering ME calc
*            W/Z + QQ production added
*            QCD jet production with qg->qg and gg->gg for onshell quark (onshell g) included
*		
*________________________________________________________________________
CASCADE 
*            Version  2.0.02-beta  (Sept 2009)
*		 meoffsh includes again correction for k^2 = kt^2 (needed for
*             reasonable description of F2 at small x
*	        bug corrected for this correction in px,py (generate phi instead of cos phi)
*            only one library created, including all files (16.Aug 2008)
*            IPRO=15 (qg ->qg ) added using valence quarks (12.Aug 2008)
*            CCFMSTFQ for the unintegrated valence quark distribution added(12.Aug 2008)
*            PTCU is now applied always in the lab frame, no longer in the CM
*             frame, since divergencies could be still there.(12.Aug 2008)
*            autotools now also with shared libraries (20. Mar 2008)
*            updated steering files, also for GENSER.(20. Mar 2008)
*            bug in event corrected: only events with xsec>0 accepted (5.4.08)
*		 NOW CERNLIB free version
* 		 in caps: for ILHA=10 MSTJ(41)=1 enforced to avoid isolated photons
*              from FPS in pythia event record
*		 Simple MC integration and generation routine included via IINT=1
*		 Valence quark distributions and g^*q -> gq and g^*g -> gg processes
		 are included.
*________________________________________________________________________
CASCADE 
*            Version  2.0.01  (24.Dec 2007)
*            improvements done on installation using autotools
*            working now with make and make install
*________________________________________________________________________
CASCADE 
*            Version  2.0.00 
*            LHA interface for PYTHIA/HERWIG included
*            ILHA=10 uses PYTHIA for final state PS and remnant treatment
*            updates in caupinit,caupevnt
*            update in p_semih.F: removed line with P(2,4)=abs(p(2,3)
*            update in caps.F restored event record also for LST(21)=55
*                             caused energy-mom mismatch before.
*                             for ILHA>1 also use caremn.F
*                             more precise energy-mom check
*            update in caremn.F for ILHA>1 set pt diquark=0
*                             for ILHA>1 limit chi<0.8
*            meoffsh: scale PT2Q changed to average of 2 outgoing quarks
*                     reduces xsection by ca 20% (compared to vers 1.2)
*________________________________________________________________________
CASCADE 
*            Version  12010
*          - insq set to 0, as for updfs... (in casbran.F)
*          - Qlam0 set to Qg0, as in updf .. (in casbran.F) 
*          - changed upper limit of xsi_hard to min(xsi_hard(2),5.d8) in cascps.F 
*                  showed up in wrong Qmax distribution for PS.
*          - no IPS for QG21 (QG22) < Q0 (in cascps.F)
*          - final state PS was always essentially switched off by IFINAL=0 in
*            caps.F. changed to default IFINAL=1.
*          - ordering in casbran.F was wrong, always only q ordering instead of
*            angular ordering (using now casbran-v24.F).
*________________________________________________________________________
CASCADE 
*            Version  12009
*            meoffsh updated: now 2 scales for pp
*            unnecessary cut in caps removed (xgtest), which caused asymmetry 
*              in parton showering
*            bug in cascps.F corrected: xsi_hard was wrong for beam 1
*________________________________________________________________________
CASCADE with LHA interface
*            Version  12008
*            ILHA = 1 added: event record written to file in LHA format to be
*                            read in by PYTHIA or HERWIG
*            ILHA = 10 added: PYTHIA (using LHA interface) is called directly to
*                            do final state PS and hadronsiation. 
*________________________________________________________________________
CASCADE 
*            Version  12007
*            date: 2004/12/23
*            bug in cascade for new gluon dat files corrected:  
*            IPS was switched off
*      
*            bug in steer corrected Nmax now at 1000 
*
*________________________________________________________________________
CASCADE (with PYTHIA6.226)
*            Version  12003
*            bug in cascps corrected 
*            date: 2004/11/09 13:42:21;  
*            steering files updated to new frag. parameters 
*            date: 2004/09/10 17:20:26; 
*            updated to read scale parameters from gluon file
*            date: 2004/11/08 06:47:41; 
*            bugs in cascps and casbran for ppbar corrected
*            bug in casbran (neg t in log) corrected (found by  Eduardo Rodrigues )
*             
*________________________________________________________________________
CASCADE  
*            Version  12000
*            Higgs production included
*            e+e- option and resolved photons included
*________________________________________________________________________
CASCADE  
*            Version  10000
*            published version for ep and pp 
*________________________________________________________________________
\end{verbatim}
\end{tiny}
\subsection{Subroutines and functions}

The source code of \CASCADE\ and this manual can be found under:\\
\verb+http://www.desy.de/~jung/cascade/+

\begin{defl}{123456789012345}
\item[{\tt CAMAIN}]
                  main program.
\item[{\tt CASINI}] 
                   to initialises the program.
\item[{\tt CASCADE}]
     to perform integration of the cross section. This routine has to be
            called before event generation can start.
\item[{\tt CAEND }] 
           to print the cross section and the number of events.
\item[{\tt CAUNIGLU(KF,X,KT,P,XPQ) }]   
      to extract the unintegrated gluon density 
	$x {\cal A}(x,k_{t},\Pmax)$ for a proton with \verb+KF=2212+,
	as a function of $x=$\verb+X+, $k_{t}^2=$\verb+KT+ and $\Pmax=$\verb+P+.
      The gluon density is returned in \verb+XPQ(0)+, where \verb+XPQ+ is an 
      array with \verb+XPQ(-6:6)+.
\item[{\tt EVENT }] 
        to perform the event generation.
\item[{\tt ALPHAS(RQ)}] 
         to give $\alpha_s (\mu)$ with $\mu = $\verb+RQ+.
\item[{\tt PARTI }] 
         to give initial particle and parton momenta.
\item[{\tt FXN1 }] 
         to call routines for selected processes:
          \verb"XSEC1".
\item[{\tt CUTG(IPRO) }] 
          to cut on $p_t$ for $2 \to 2$ process 
          in integration and event generation.
\item[{\tt MEOFFSH }] 
          matrix element for 
	    $\gamma^* g^* \rightarrow q \bar{q}$  and
	    $g^* g^* \rightarrow q \bar{q}$ 
	    including quark masses. $q$ can be light or heavy quarks.
\item[{\tt MEHIGGS }] 
          matrix element for 
	    $\gamma^* g^* \rightarrow h^0$.
\item[{\tt DOT(A,B) }] 
         four-vector dot product of $A$ and $B$.
\item[{\tt DOT1(I,J)}]
          four-vector dot product of vectors I and J in
          \verb"PYJETS" common.
\item[{\tt PHASE }] 
         to generate momenta of final
         partons in a $2 \rightarrow 2$ subprocess according to phase space.	   
\item[{\tt P\_SEMIH }] 
         to generate kinematics and the 
         event record for $ep$, $\gamma p$ and $p\bar{p}$
         processes.
\item[{\tt CAREMN(IPU1,IPU2) }]   
         to generate the beam remnants.
            Copied from LEPTO 6.1~\cite{\LEPTOMC} and updated for
            the use in \CASCADE . 
\item[{\tt CASPLI(KF,KPA,KFSP,KFCH) }]   
            to give the spectator \verb"KFSP" and \verb"KFCH"
            partons when a parton
            \verb"KPA" is removed from particle \verb"KF".
            Copied from LEPTO 6.1~\cite{\LEPTOMC} and updated for
            the use \CASCADE. 
\item[{\tt CAPS }]   
  to generate color flow  for all
            processes and prepare for initial and final state
            parton showers.
\item[{\tt CASCPS(IPU1,IPU2) }]   to generate initial state radiation.
\item[{\tt COLORFLOW }]   to generate color configuration for $g^*g \to gg$ and $g^*q
            \to qg$ processes.
\item[{\tt GADAP }]   Gaussian integration routine for 1-dim and 
            2-dim integration.
            Copied from LEPTO 6.1~\cite{\LEPTOMC}.
\end{defl}

\subsection{Parameter switches}
\begin{defl}{123456789012345}
\item[]   BASES/SPRING integration procedure.
\item[{\tt NCAL:}]  (D:=20000) Nr of calls per iteration for bases.
\item[{\tt ACC1:}]  (D:=1)    relative precision (in \%) for grid optimisation.
\item[{\tt ACC2:}]  (D:=0.5)  relative precision (in \%) for integration.
\item[]   Event record output.
\item[{\tt ILHA:}] \index{ILHA} (D: = 0) output in LHA accord \cite{lhaaccord} format
\item[ ]  = 0: \CASCADE\ type output of event record
\item[ ]  = 1: output of event record to be read in by fragmentation programs	
\item[ ]  = 10:  use LHA format to produce remnant and fragmentation in \PYTHIA\ style.	

\end{defl}

\subsubsection{Parameters for kinematics}
\begin{defl}{123456789012345}
\item[{\tt PBE1:}] \index{PBE1}
                     (D:=$-30$)   momentum $p$ [GeV/$c$]
                        of incoming hadron 1 (\verb"/INPU/").
\item[{\tt KBE1:}] \index{KBE1} Lund flavour code of incoming hadron 1 (\verb+KBE1=11+
for electrons, \verb+KBE1=22+ for photons, \verb+KBE1=2212+ for protons )
\item[{\tt IRE1:}] \index{IRE1}  hadron/lepton 1 has a structure (\verb+IRE1=1+) or
interacts directly with the target (\verb+IRE1=0+ for a DIS electron) 
\item[{\tt PBE2:}] \index{PIN}
                     (D:=$820$    momentum $p$ [GeV/$c$]
                        of incoming proton (\verb"/INPU/").
\item[{\tt KBE2:}] \index{KBE2} Lund flavour code of incoming hadron 2 (\verb+KBE2=11+
for electrons, \verb+KBE2=22+ for photons, \verb+KBE2=2212+ for protons )
\item[{\tt IRE2:}] \index{IRE2}  hadron/lepton 2 has a structure (\verb+IRE2=1+) or
interacts directly with the target (\verb+IRE2=0+ for a DIS electron) 
\item[{\tt NFLAV}]  \index{NFLAV}
(D: = 5) number of active flavours, can be set by user 
                    ({\tt /CALUCO/}).

\end{defl}
\subsubsection{Parameters specific for leptoproduction}
\begin{defl}{123456789012345}
\item[{\tt QMI:}] \index{QMI} 
       (D: = 5.0) (\verb"/VALUES/") minimum $Q^2$ to be generated 
\item[{\tt QMA:}] \index{QMA}
 (D: = $10^8$) (\verb"/VALUES/") maximum $Q^2$ to be generated.
\item[{\tt YMI:}] \index{YMI} 
(D: = 0.0) (\verb"/VALUES/") minimum $y$ to be generated.
\item[{\tt YMA:}] \index{YMA}  
(D: = 1.0) (\verb"/VALUES/") maximum $y$ to be generated.
\item[{\tt THEMA,THEMI}] \index{THEMA,THEMI}
  (D: {\tt THEMA} = 180., {\tt THEMI} = 0)
                    maximum and minimum scattering angle $\theta$ of the 
                    electron ({\tt /CAELEC/}).
\end{defl}

\subsubsection{Parameters for hard subprocess selection}
\begin{defl}{123456789012345}
\item[{\tt IPRO:}] \index{IPRO} (D: = 10) ({\tt /CAPAR1/})
                    selects hard subprocess to be generated. 
\item[{\it         }]
                 =2:   $\gamma g^* \rightarrow J/\psi (\Upsilon) g$, $g^* g^* \rightarrow J/\psi (\Upsilon) g$.
\item[{\it         }]
                 =3:   $g^* g^* \rightarrow \chi_{c(b)}$.
\item[{\it         }]
                 =10:   $\gamma^* g^* \rightarrow q \bar{q}$,
		            $g^* g^* \rightarrow q \bar{q}$,
                        $g^* g \to gg $ and 
                        $g^* q \to gq$ 
                        for light quarks.
\item[{\it         }]
                 =11:   $\gamma^* g^* \rightarrow Q \bar{Q}$ or 
		            $g* g^* \rightarrow Q \bar{Q}$
                        for heavy quarks.
\item[{\it         }]
                 =102:   $g^* g^* \rightarrow h^0$
                        for Higgs production in hadron-hadron collisions.
\item[{\it         }]
                 =501:   
                 $q g^* \rightarrow Z q$
                        for $Z$+jet production in hadron-hadron collisions.
 \item[{\it         }]
                 =503:   $g* g^* \rightarrow Z q\bar{q}$
                        for $Z$+jet production in hadron-hadron collisions. 
                 =504:   $g* g^* \rightarrow Z Q\bar{Q}$
                         The flavour index of the heaviest quark is selected via  \verb+IHFLA+
 \item[{\it         }]
                 =513:   $g* g^* \rightarrow W q_i q_j$
                        for $W$+jet production in hadron-hadron collisions. 
 \item[{\it         }]
                 =514:   $g* g^* \rightarrow W q_i Q_j$
                        for $W$+jet production in hadron-hadron collisions. 
\item[{\tt IRPA:}] \index{IRPA} 
			= 1   $g^* g^* \rightarrow q \bar{q}$ switched on for \verb+IPRO=10+.                    
\item[{\tt IRPB:}] \index{IRPB} 
			= 1   $g^* g \rightarrow gg$ switched on for \verb+IPRO=10+.                     
\item[{\tt IRPC:}] \index{IRPC} 
			= 1   $g^* q \rightarrow g q $ switched on for \verb+IPRO=10+.    
\item[{\tt IHFLA:}] \index{IHFLA}  (D:=4)
			= 4 flavor of heavy quark produced (in \verb+IPRO=10+,   \verb+IPRO=504+ and \verb+IPRO=514+ ).		                 
\item[{\tt IPSIPOL:}] \index{IPSIPOL}  (D:=0)
			= 1   use matrixelement including $J/\psi$ ($\Upsilon$) polarisation and subsequent leptonic decay 
			 for \verb+IPRO=2+.                     
\item[{\tt PT2CUT(IPRO):}] \index{PT2CUT}
 (D=0.0) minimum $\hat{p}^2 _{\perp}$ for
                            process {\tt IPRO} ({\tt /CAPTCUT/}).       
\end{defl}
\subsubsection{Parameters for parton shower and fragmentation}
\begin{defl}{123456789012345}
\item[{\tt NFRAG:}] \index{NFRAG} (D: = 1)
                        switch for fragmentation ({\tt /CAINPU/}).
\item[] = 0: off
\item[] = 1: on 
\item[{\tt IFPS:}] \index{IFPS} (D: = 3)
                  switch for parton shower ({\tt /CAINPU/}).
\item[] = 0: off
\item[] = 1: initial state
\item[] = 2: final state
\item[] = 3: initial and final state  
\item[{\tt ITIMSHR:}] \index{ITIMSHR} (D: =1)
\item[] =0: no shower of time like partons
\item[] =1: time like partons may shower
\item[{\tt ICCFM:}] \index{ICCFM} (D: =1)
\item[] =1: CCFM evolution (all loops)
\item[] =0: DGLAP type evolution (one loop)
\item[{\tt IFIN}] \index{IFIN} (D:=1)  scale switch for  final state parton shower
\item[] = 1: $\mu^2=2 (m^2_{1\;t} + m^2_{2\;t})$
\item[] = 2: $\mu^2=\hat{s}$
\item[] = 3:  $\mu^2=2 (m^2_{1} + m^2_{2})$
\item[{\tt SCAF}]\index{SCAF} (D:=1.) scale factor for final state parton shower
\end{defl}

\subsubsection{Parameters for structure functions, $\alpha_s$ and scales}
\label{sec:alphas}
\begin{defl}{123456789012345}
\item[{\tt IRUNAEM:}] \index{IRUNAEM} (D: = 0) ({\tt /CAPAR1/})
                   select running of $\alpha _{em}(Q^2)$.
\item[]
                        =0:  no running of $\alpha _{em}(Q^2)$
\item[]
                        =1:  running of $\alpha _{em}(Q^2)$

\item[{\tt IRUNA:}] \index{IRUNA} (D: = 1)
                        switch for running $\alpha _s$.
\item[]
                        =0:  fixed $\alpha_s=0.3$ 
\item[]
                        =1: running $\alpha _s(\mu^2)$
\item[{\tt IQ2:}] \index{IQ2} (D: = 3)
                   select scale $\mu^2$ for $\alpha _s(\mu^2)$.
\item[]
                        =1:  $\mu^2 = 4 \cdot m_{q} ^2$
                             (use only for heavy quarks!)
\item[]
                        =2:  $\mu^2 = \hat{s} $
                             (use only for heavy quarks!)
\item[]
                        =3:  $\mu^2 = 4 \cdot m^2 + p_{\perp} ^2$
                             
\item[]
                        =4:  $\mu^2 = Q^2$
                        
\item[]
                        =5:  $\mu^2 = Q^2 + p_{\perp} ^2 + 4 \cdot m^2 $
\item[{\tt IGLU:}] \index{IGLU} (D: = 1010)
			select unintegrated gluon density ({\tt /GLUON/}).
\item[]	Note that initial state parton showers not possible for 
\verb+IGLU=2, 3, 4, 5+
\item[]
                        =1:  CCFM old set JS2001 \cite{jung_salam_2000}	
\item[]                 =  1001: CCFM J2003 set 1 \cite{jung-dis03}
\item[]                 =  1002: CCFM J2003 set 2 \cite{jung-dis03}
\item[]                 =  1003: CCFM J2003 set 3 \cite{jung-dis03}
\item[]                 =  1010: CCFM set A0 \cite{jung-dis04}
\item[]                 =  1011: CCFM set A0+ \cite{jung-dis04}
\item[]                 =  1012: CCFM set A0- \cite{jung-dis04}
\item[]                 =  1013: CCFM set A1 \cite{jung-dis04}
\item[]                 =  1020: CCFM set B0 \cite{jung-dis04}
\item[]                 =  1021: CCFM set B0+ \cite{jung-dis04}
\item[]                 =  1022: CCFM set B0- \cite{jung-dis04}
\item[]                 =  1023: CCFM set B1 \cite{jung-dis04}
\item[]                 =  1101: CCFM set C \cite{jung-dis07}
\item[]
                        =2:  derivative of GRV~\cite{GRV95}
				$\frac{d xg(x,Q^2)}{dQ^2}$.
\item[]
                        =3:  approach of Bl\"umlein~\cite{Bluemlein}
\item[]
                        =4:  KMS~\cite{martin_stasto} (\verb+kms.dat+)
\item[]
                        =5:  saturation model~\cite{wuesthoff_golec-biernat}
\item[]
                        =6:  KMR~\cite{martin_kimber} (\verb+kmr.dat+)

\end{defl}

\subsubsection{Accessing information}
\begin{defl}{123456789012345}
\item[ ]

\item[{\tt AVGI}] \index{AVGI} integrated cross section ({\tt /CAEFFIC/}).
\item[{\tt SD}] \index{SD} standard deviation of integrated cross section
                 ({\tt /CAEFFIC/}).
\item[ ]
\item[{\tt SSS}] \index{SSS} 
squared center of mass energy  $s$ ({\tt /CAPARTON/}).
\item[{\tt PBEAM}] \index{PBEAM} energy momentum vector of beam particles
  ({\tt /CABEAM/}).
\item[{\tt KBEAM}] \index{KBEAM} flavour code of beam particles
 ({\tt /CABEAM/}).
\item[{\tt Q2}] \index{Q2} in leptoproduction: actual $Q^2$ of
                         $\gamma$ ({\tt /CAPAR4/}).
\item[{\tt YY}] \index{YY}
negative light-cone momentum
                       fraction of parton $1$ ($\gamma^*$, $g^*$)
				({\tt /CASGKI/}).
\item[{\tt YY\_BAR}] \index{YY\_BAR}
positive light-cone momentum
                       fraction parton $1$ ($\gamma^*$, $g^*$)
				({\tt /CASGKI/}).
\item[{\tt XG}] \index{XG}
positive light-cone momentum
                        fraction of parton $2$ ($g^*$) 
				({\tt /CASGKI/}).
\item[{\tt XG\_BAR}] \index{XG\_BAR}
negative light-cone momentum
                        fraction of parton $2$ ($g^*$)
				({\tt /CASGKI/}).
\item[{\tt KT2\_1,KT2\_2}] \index{KT2\_1,KT2\_2} transverse momenta squared
			$k_{t\;1(2)}^2$ [GeV$^2$] of partons 
 			$1(2)$ which enter to the matrix element.
\item[{\tt YMAX,YMIN}] \index{YMAX,YMIN} actual upper and lower limits for
			$y=$\verb+YY+ 
                          ({\tt /CAPAR5/}).
\item[{\tt Q2MAX,Q2MIN}] \index{Q2MAX,Q2MIN} actual upper and
                        lower limits for $Q^2$ (corresponding to \verb+KT2_1+)
				 of $\gamma$ ({\tt /CAPAR5/}).
\item[{\tt XMAX,XMIN}] \index{XMAX,XMIN} 
upper and lower limits for $x$ ({\tt /CAPAR5/}).

\item[{\tt AM(18)}] \index{AM}
 vector of masses of final state particles of hard
                         interaction ({\tt /CAPAR3/}).
\item[{\tt SHAT}] \index{SHAT}
                        invariant mass $\hat{s}$ [GeV$^2$]
                     of hard subprocess ({\tt /CAPAR5/}).

\item[{\tt NIA1,NIA2}] \index{NIA1,NIA2}
 position of partons in hard interaction in
                         {\tt PYJETS} event record ({\tt /CAHARD/}).
\item[{\tt NF1,NF2}] \index{NF1,NF2} first and last position final
                         partons/particles of
                         hard interaction in {\tt PYJETS} ({\tt /CAHARD/}).
\item[{\tt Q2Q}] \index{Q2Q} hard scattering scale $\mu ^2$ used in
                         $\alpha_s$ and structure functions ({\tt /CAPAR4/}).
\item[{\tt ALPHS}] \index{ALPHS} actual $\alpha_s$ ({\tt /CAPAR2/}).
\item[{\tt ALPH}] \index{ALPH} $\alpha_{em}$ ({\tt /CAPAR2/}).
\item[{\tt NIN}] \index{NIN} 
number of trials for event generation ({\tt /CAEFFIC/}).
\item[{\tt NOUT}] \index{NOUT} number of successful generated events 
({\tt /CAEFFIC/}).

\end{defl}

\subsection{List of COMMON blocks}
\verb"  COMMON/CABEAM/PBEAM(2,5),KBEAM(2,5),KINT(2,5)" \\
\verb"  COMMON/CAHARD/NIA1,NIA2,NIR2,NF1,NF2"\\
\verb"  COMMON/CAHFLAV/IHFLA"\\
\verb"  COMMON/CAINPU/PLEPIN,PPIN,NFRAG,ILEPTO,IFPS,IHF,INTER,ISEMIH"\\
\verb"  COMMON/CALUCO/KE,KP,KEB,KPH,KGL,KPA,NFLAV"\\
\verb"  COMMON/CAEFFIC/AVGI,SD,NIN,NOUT"\\
\verb"  COMMON/CAELEC/THEMA,THEMI"\\
\verb"  COMMON/CAGLUON/IGLU"\\
\verb"  COMMON/CAPAR1/IPRO,IRUNA,IQ2,IRUNAEM"\\
\verb"  COMMON/CAPAR2/ALPHS,PI,ALPH,IWEI"\\
\verb"  COMMON/CAPAR3/AM(18),PCM(4,18)"\\
\verb"  COMMON/CAPAR4/Q2,Q2Q"\\
\verb"  COMMON/CAPAR5/SHAT,YMAX,YMIN,Q2MAX,Q2MIN,XMAX,XMIN"\\
\verb"  COMMON/CAPAR6/LST(30),IRES(2)"\\
\verb"  COMMON/CAPARTON/SSS,CM(4),DBCMS(4)"\\
\verb"  COMMON/CAPTCUT/PT2CUT(20)"\\
\verb"  COMMON/CASKIN/YY,YY_BAR,XG,XG_BAR,KT2_1,KT2_2,PT2H,SHH"\\
\verb"  COMMON/VALUES/QMI,YMI,QMA,YMA"\\
\section{Example Program}
\begin{tiny}
\begin{verbatim}

      PROGRAM CASMAIN
      Implicit None
      Integer N1,N2
      DOUBLE PRECISION  PLEPIN,PPIN
      INTEGER KE,KP,KEB,KPH,KGL,KPA,NFRAG,ILEPTO,IFPS,IHF
      INTEGER INTER,ISEMIH
      INTEGER NIA1,NIR1,NIA2,NIR2,NF1,NF2,NFT,NFLAV
      COMMON/CALUCO/KE,KP,KEB,KPH,KGL,KPA,NFLAV
      COMMON/CAINPU/PLEPIN,PPIN,NFRAG,ILEPTO,IFPS,IHF,INTER,ISEMIH
      COMMON/CAHARD/NIA1,NIA2,NIR2,NF1,NF2
      INTEGER IHFLA
      COMMON/CAHFLAV/IHFLA

      DOUBLE PRECISION THEMA,THEMI,PT2CUT
      INTEGER IRUNA,IQ2,IRUNAEM
      INTEGER IPRO
      COMMON/CAPAR1/IPRO,IRUNA,IQ2,IRUNAEM
      COMMON/CAELEC/ THEMA,THEMI
      COMMON/CAPTCUT/PT2CUT(20)
      REAL ULALPS,ULALEM
      EXTERNAL ULALPS,ULALEM
      DOUBLE PRECISION QMI,YMI,QMA,YMA
      COMMON/VALUES/QMI,YMI,QMA,YMA

      Integer Iglu
      Common/CAGLUON/Iglu
	
      Integer ISEED,I
	
	
      ISEED = 124567
      n1=0
      n2=0
C initialize random number generator	
      CALL RM48IN(ISEED,N1,N2)
C initialize PYTHIA 6 parameters
      CALL GPYINI
C initialize CASCADE parameters
      CALL CASINI


C Select parton shower (IPS=1 initial, =2 final, 3 initial+final PS )
      IFPS = 3
C scale for alpha_s
C IQ2 =1 mu^2 = 4m_q^2 (m_q = light quark or heavy quark depending on IPRO)
C IQ2 =2 mu^2  = shat
C IQ2 =3 mu^2  = 4m_q^2 + pt^2 (m_q = light quark or heavy quark depending on IPRO)
C IQ2 =4 mu^2  = q^2 (q^2 of virtual photon)
C IQ2 =5 mu^2  = q^2 + pt^2 + 4m_q^2(q^2 of virtual photon)
      IQ2=3
C select process (IPRO=10 for light quarks, IPRO=11 for heavy quarks)
      IPRO= 10
C total number of flavours involved
      NFLAV = 4
C select unintegrated gluon density (D=1)
      Iglu = 1
C minimum Q^2 of electron to be generated
      QMI = 0.5d0
C maximum Q^2 of electron to be generated
      QMA = 10D8
C minimum y of electron to be generated
      YMI=0.0d0
C minimum y of electron to be generated
      YMA=1.0d0
C maximum theta angle of scattered electron
      THEMA = 180.0D0
C minimum  theta angle of scattered electron
      THEMI =   0.0D0
C momentum of beam 1 (electron,proton,antiproton)
      PLEPIN =-27.5
C Lund flavour code for beam 1 (electron=11,photon=22,proton=2212,antiproton=-2212)
      KE=11
C momentum of beam 2 (proton)
      PPIN   = 820.
C perform fragmentation NFRAG=0/1
      NFRAG = 1
c for IPRO = 11  which flavour is produced
      IHFLA = 4
c 
c Start integration of x-section
c 
      CALL CASCADE
c 
c Print out result of integration of x-section
c 
      CALL CAEND(1)

c 
c Start event loop
c 
      Do I=1,100
c generate an event
         CALL EVENT
      Enddo
c 
c Print out of generated events summary
c 
      CALL CAEND(20)

      STOP
      END
             
\end{verbatim}
\end{tiny}
\section{Program Installation}
\CASCADE\ now follows the standard AUTOMAKE
convention. To install the program, do the following
\begin{verbatim}
1) Get the source

tar xvfz cascade-XXXX.tar.gz
cd cascade-XXXX

2) Set environment variables for PYTHIA
example (Please change to the proper path of the libraries):
in csh: 
setenv PYTHIA "/home/jung/cvs/pythia6422"

in zsh:
export PYTHIA="/home/jung/cvs/pythia6422"

2) Generate the Makefiles
./configure --prefix=install=path --disable-shared

3) Compile the binary
make

4) Install the executable and PDF files
make install 

4) The executable is in bin
set the path for the updf data files, if different from 
the default (for example)

export PDFPATH=/Users/jung/jung/cvs/cascade2/cascade-2.2.0/share

run it with:
cascade < steer_pp-bottom

\end{verbatim}
\section{Acknowledgments}
We are very grateful to B.~Webber for providing us with the \SMALLX~
code, which was the basis for the \CASCADE\ Monte Carlo generator.  
We are very grateful also to G.~Ingelman and T.~Sj\"ostrand for many discussions and for their courtesy to let us use their code for proton remnant treatment.
One of us (H.J.) enjoyed very much the collaboration with G.~Salam and his patience and help in all
different kinds of discussions concerning CCFM and a backward
evolution approach.
We have enjoyed and learned a lot from the discussions with
B.~Andersson, G.~Gustafson, L.~J\"onsson, H.~Kharraziha
and L.~L\"onnblad during several years.  
Some of us (S.B., H.J., A.L and N.Z)  are very grateful to 
DESY Directorate for the support in the 
framework of Moscow -- DESY project on Monte-Carlo
implementation for HERA -- LHC.

\begin{theindex}

  \item ALPH, 14
  \item ALPHS, 14
  \item AM, 14
  \item AVGI, 14

  \indexspace

  \item ICCFM, 13
  \item IFPS, 13
  \item IGLU, 13
  \item IPRO, 12
  \item IQ2, 13
  \item IRE1, 12
  \item IRE2, 12
  \item IRPA, 12
  \item IRPB, 12
  \item IRPC, 12
  \item IRUNA, 13
  \item IRUNAEM, 13

  \indexspace

  \item KBE1, 12
  \item KBE2, 12
  \item KBEAM, 14
  \item KT2\_1,KT2\_2, 14

  \indexspace

  \item NF1,NF2, 14
  \item NFLAV, 12
  \item NFRAG, 13
  \item NIA1,NIA2, 14
  \item NIN, 14
  \item NOUT, 14

  \indexspace

  \item PBE1, 12
  \item PBEAM, 14
  \item PIN, 12
  \item PT2CUT, 13

  \indexspace

  \item Q2, 14
  \item Q2MAX,Q2MIN, 14
  \item Q2Q, 14
  \item QMA, 12
  \item QMI, 12

  \indexspace

  \item SD, 14
  \item SHAT, 14
  \item SSS, 14

  \indexspace

  \item THEMA,THEMI, 12

  \indexspace

  \item XG, 14
  \item XG\_BAR, 14
  \item XMAX,XMIN, 14

  \indexspace

  \item YMA, 12
  \item YMAX,YMIN, 14
  \item YMI, 12
  \item YY, 14
  \item YY\_BAR, 14

\end{theindex}

\bibliographystyle{heralhc} 
\raggedright 
\bibliography{cascade}

\end{document}